# Robust light bullets in Rydberg gases with moiré lattice

Ze-Yang Li[1], Jun-Hao Li[1,2], Yuan Zhao[2,3], Jin-Long Cui[1,2], Jun-Rong He[1,2], Guo-Long Ruan[1]*, Boris A. Malomed[4,5], and Si-Liu Xu[2,3]*

[1] *School of Electronics and Information Engineering, Hubei University of Science and Technology, Xianning 437100, China*
[2] *Laboratory of Optoelectronic Information and Intelligent Control, Hubei University of Science and Technology, Xianning 437100, China*
[3] *School of Biomedical Engineering and Medical Imaging, Xianning Medical College, Hubei University of Science and Technology, Xianning 437100, China*
[4] *Department of Physical Electronics, School of Electrical Engineering, Faculty of Engineering, Tel Aviv University, P.O.B. 39040, Ramat Aviv, Tel Aviv, Israel*
[5] *Instituto de Alta Investigación, Universidad de Tarapacá, Casilla 7D, Arica, Chile*
*\* Corresponding author: ruangl@126.com (Guo-Long Ruan); xusiliu1968@163.com (Siliu Xu)*

**Abstract:** Rydberg electromagnetically-induced transparency has been widely studied as a medium supporting light propagation under the action of nonlocal nonlinearities. Recently, optical potentials based on moiré lattices (MLs) were introduced for exploring unconventional physical states. Here, we predict a possibility of creating fully three-dimensional (3D) light bullets (LBs) in cold Rydberg gases under the action of ML potentials. The nonlinearity includes local self-defocusing and long-range focusing terms, the latter one induced by the Rydberg-Rydberg interaction. We produce zero-vorticity LB families of the fundamental, dipole, and quadrupole types, as well as vortex LBs. They all are gap solitons populating finite bandgaps of the underlying ML spectrum. Stable subfamilies are identified utilizing the combination of the anti-Vakhitov-Kolokolov criterion, computation of eigenvalues for small perturbations, and direct simulations.

## 1. Introduction

Light bullets (LBs) are two- and three-dimensional (2D and 3D) spatiotemporal solitons that emerge in nonlinear optical media [1]. Studies of LBs have drawn much interest due to rich phenomenology and potential applications of these objects [2-6]. Experimental signatures of LBs have been reported in different types of optical media, including quasi-2D spatiotemporal solitons supported by the second-harmonic-generating nonlinearity [7], transient 3D semi-discrete LBs in waveguide arrays [8], 3D pulse-train solitons [9], and LBs in multimode fibers [10].

However, 3D LBs are generally subject to strong instabilities, driven by the supercritical wave collapse in the 3D space and, in addition, spontaneous splitting of solitons with embedded vorticity [3,6,11]. Two-dimensional LBs, which may exist as self-trapped spatiotemporal modes in planar optical waveguides are vulnerable to weaker instability, driven by the critical collapse [3,6]. Several setups have been proposed to generate stable 3D and 2D multidimensional solitons in optics, such as spatially patterned nonlinear interactions [5], optical tandem systems [12], materials with saturable, competing or nonlocal nonlinearities [13-15], *PT* (parity-time) -symmetric photonic structures [16,17], waveguiding lattices [18], and electromagnetically induced transparency (EIT) in cold Rydberg atomic gases [19,20].

Cold atomic media offer an ideal setting for the creation of stable multidimensional solitons, both zero-vorticity (fundamental) and vortical ones [6,11]. In this context, cold Rydberg atoms can be used for this purpose, owing to the strong long-range Rydberg-Rydberg interaction (RRI) between remote atoms and their long lifetimes [21,22]. In particular, the RRI can be mapped into nonlocal optical nonlinearity through EIT, which is a strong effect even at the single-photon level [21]. As a result, various optical phenomena based on the RRI have been predicted, such as storage and retrieval of stable LBs [19], two-component ones [23], controllable bound states of LBs [24], transient optical response [25], and effective photon interactions [26]. Long lifetimes of Rydberg atomic states, on the order of tens of microseconds, secure the coherence of the induced optical nonlinearities. These properties offer a nearly ideal environment for the design of new photonic devices, such as single-photon switches and transistors [27,28], quantum memory, and phase gates [29,30].

Various external potentials, such as Bessel lattices [31] and *PT*-symmetric ones [32], applied to Rydberg gases, may be used for generating stable LBs in them. Recently, much interest has been drawn to optical moiré lattices (MLs) [33-37]. They are built as 2D twisted superpositions of two identical periodic structures overlapping at an angle. MLs may induce both periodic and aperiodic potentials in the atomic gas, naturally providing new options for the study of the LB dynamics in these potentials, in comparison to the previously studied ones. In particular, ML potentials, acting on the atomic wave function, induce bandgaps in their spectra specific to the shape of these potentials, which may provide a habitat for self-trapped states in the form of gap solitons (GSs) [6,38-40].

The present paper aims to construct stable three-dimensional LBs in Rydberg gases with optical MLs, utilizing a systematic numerical analysis. The results reveal families of LBs, including fundamental, dipole, quadruple, and vortex 3D solitons. Crucially important stability conditions are obtained for them, depending on parameters of nonlocal RRI and ML potentials. Compared to the 3D LBs in Rydberg gases system[19], this work introduced the moiré lattice potential and the GSs were obtained. On the other hand, in ref.[20], the authors addressed 2D solitons in the complex PT-symmetric potential, while we here consider the 3D setting, with the real potential.

## 2. The model

A Rydberg-dressed four-level cold atomic system is proposed here, as shown in Fig. 1(a). A standard $\Lambda$-type EIT structure is constructed by states $|1\rangle$, $|2\rangle$, $|3\rangle$, the probe, and control laser fields. The pulsed probe (with frequency $\omega_p$, half Rabi frequenc $\Omega_p$) and control (with frequency $\omega_c$, half Rabi frequency $\Omega_c$) laser fields are coupled to the transitions $|1\rangle \to |3\rangle$ and $|2\rangle \to |3\rangle$, where $\Gamma_{ij}$ are spontaneous emission decay rates between states $|i\rangle$ and $|j\rangle$. This EIT structure is dressed by a high-lying Rydberg state $|4\rangle$ which is coupled by the auxiliary field (with frequency $\omega_a$, half Rabi frequency $\Omega_a$). The interaction between Rydberg atoms is described by the van der Waals potential $V_{\text{vdw}} = \hbar V(\mathbf{r}' - \mathbf{r})$, with $V(\mathbf{r}' - \mathbf{r}) = C_6 / (R_b^6 + |\mathbf{r}' - \mathbf{r}|^6)$, $C_6$ is the dispersion parameter and $R_b$ is the Rydberg blockade radius, $\mathbf{r}$ and $\mathbf{r}'$ are position vectors of Rydberg atoms.

Suppose the external potential is the ML-induced potential. To obtain the electromagnetically induced moiré optical lattices under the EIT condition, the control field is chosen as a periodic function of spatial coordinates $(x, y)$ [33], i.e.,

$$\Omega_c(x, y) = \Omega_{c0}[1 + f(x, y)], \tag{1}$$

where $\Omega_{c0}$ is the magnitude of the control field, and the term $f(x,y) = V_1(\cos^2 x + \cos^2 y) + V_2(\cos^2 x' + \cos^2 y')$ is a periodic function of spatial coordinates, where $V_{1,2} > 0$ are modulation depths of the two sublattices characterized by coordinates $(x, y)$ and $(x', y')$. The rotated coordinates $(x', y')$ are defined through the twist angle $\theta$

$$\begin{pmatrix} x' \\ y' \end{pmatrix} = \begin{pmatrix} \cos\theta & -\sin\theta \\ \sin\theta & \cos\theta \end{pmatrix} \begin{pmatrix} x \\ y \end{pmatrix}. \tag{2}$$

The ML structure can be transformed from periodic to aperiodic ones by changing the twist angle. Only with finite $\theta$, such as $\theta = \arctan(4/3)$ and $\theta = \arctan(12/5)$, the ML shows periodic structures [34]. The probe field is supposed to be weak, hence $\Omega_p \ll \Omega_c$ is taken as a small quantity. Thus, the potential induced by the optical ML takes the non-dimensional form[33]

$$V_{\text{OL}}(x, y) = -\frac{V_0}{[1 + f(x, y)]^2}, \tag{3}$$

where $V_0$ is the effective lattice depth. Here, the periodic function $f(x, y)$ can be used to manipulate the control field.

The ML potential is displayed in Fig. 1(b). For the ML with $\theta = \arctan(4/3)$ the propagation constant and the first reduced Brillouin zone in the 2D reciprocal space are shown in Figs. 1(c) and (d), respectively. The gaps are shown, vs. $p = V_1/V_2$ and lattice depth $V_0$, in Figs. 1(e) and (f). When $p < 1.5$ there exist some cross-band structures. The bands are compressed tightly and no band gaps exist when the lattice potential is absent ($V_0 = 0$). With the increase of $V_0$, the first and second finite bandgaps become wider. The ML-induced bandgap spectrum can be readily tuned utilizing the twist angle and modulation depth, which, in turn, suggests possibilities to control the formation of GSs under the action of the defocusing nonlinearity. In this work, simulations of GSs are performed at $V_0 = 1$.

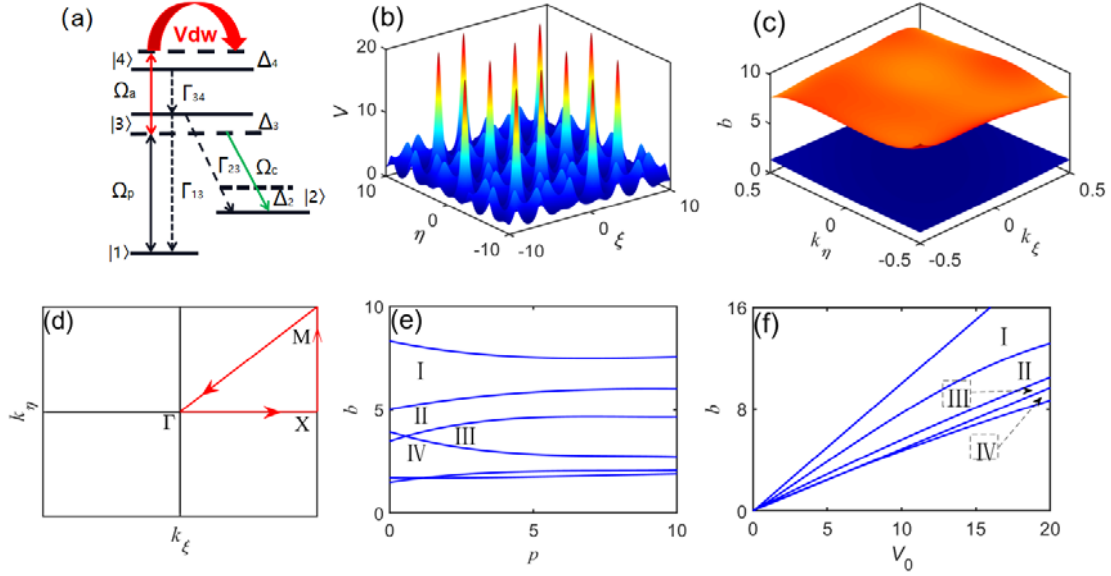

Fig. 1 (a) Rydberg-dressed four-level atomic system. (b) The potential induced by the optical ML in the $\xi$ and $\eta$ directions, where $(\xi,\eta)=(x,y)/R_\perp$ with $R_\perp$ is the typical radius of the probe pulse. (c,d) The propagation constant and first Brillouin zone in the 2D reciprocal space, where $\Gamma$, X, and M are defined as $(0,0,0)$, $(1,0,0)$, and $(1,1,0)$, respectively. (e,f) The bandgap structure is displayed with respect to ratio $p=V_1/V_2$ and lattice depth $V_0$, respectively. I, II, III, and IV in panels (e) and (f) are numbers of bandgaps between the nearest bands. The twist angle of the ML is $\theta=\arctan(4/3)$, and $V_2=1$.

The dynamics of this 4-level Rydberg atomic system is described by the Hamiltonian, $\hat{H}_H(t)=N_a\int_{-\infty}^{+\infty}d^3\mathbf{r}\hat{H}_H(\mathbf{r},t)$, where $N_a$ is the mean atomic density, and $\hat{H}_H(\mathbf{r},t)$ is Hamiltonian density with position vector $\mathbf{r}$. With the help of the electric-dipole and rotating-wave approximations, the Hamiltonian density can be written as

$$\hat{H}_H(\mathbf{r},t)=\sum_{j=1}^{4}\hbar\Delta_j\hat{S}_{13}(\mathbf{r},t)-\hbar[\Omega_p\hat{S}_{13}(\mathbf{r},t)+\Omega_a\hat{S}_{34}(\mathbf{r},t)+\Omega_c\hat{S}_{23}(\mathbf{r},t)+H.c.]$$
$$+N_a\int d^3\mathbf{r}'\hat{S}_{44}(\mathbf{r}',t)\hbar V(\mathbf{r}'-\mathbf{r})\hat{S}_{44}(\mathbf{r},t) \quad (4)$$

Here, $\hat{S}_{jl}=|l\rangle\langle j|\exp i[(\mathbf{k}_l-\mathbf{k}_j)\cdot\mathbf{r}-(\omega_l-\omega_j+\Delta_l-\Delta_j)t]$ is the transition operator

between states $|j\rangle$ and $|l\rangle$, $\Delta_3 = (\omega_3 - \omega_1) - \omega_p$ is the one-photon detuning, $\Delta_2 = \omega_p - \omega_c - (\omega_2 - \omega_1)$ and $\Delta_4 = \omega_p + \omega_a - (\omega_4 - \omega_1)$ are two-photon detunings, where $\omega_{p/c/a}$ and $\omega_{1/2/3/4}$ are frequencies of the probe/control/auxiliary laser fields and states $|1\rangle/|2\rangle/|3\rangle/|4\rangle$. Then the density-matrix $\hat{\rho}$ is obtained by solving the Bloch equations,

$$\partial\hat{\rho}/\partial t = -i\left[\hat{H}, \hat{\rho}\right]/\hbar - \Gamma[\hat{\rho}], \tag{5}$$

where $\rho_{jl} \equiv \langle \hat{S}_{jl} \rangle$ is the density elements, $\Gamma$ is the relaxation-rate matrix, and $\hat{H}$ is the Hamiltonian of this Rydberg-dressed 4-level atomic system[21].

The probe field takes the form $\Omega_p \sim F\exp[i(Kz - \omega t)]$ and satisfies the following Maxwell's equation under the slowly varying envelope approximation

$$i\left(\frac{\partial}{\partial z} + \frac{1}{c}\frac{\partial}{\partial t}\right)\Omega_p + \frac{c}{2\omega_p}\left(\frac{\partial^2}{\partial x^2} + \frac{\partial^2}{\partial y^2}\right)\Omega_p + \frac{\omega_p}{2c}\chi_p\Omega_p = 0 \tag{6}$$

where the terms with the second derivatives represent, as usual, the paraxial diffraction. Here $\chi_p = \chi_p^{(1)} + \chi_{p,1}^{(3)}|\Omega_p|^2 + \int d^3\mathbf{r}'\chi_{p,2}^{(3)}(\mathbf{r}'-\mathbf{r})|\Omega_p|^2$ is the effect susceptibility for the probe field with third-order approximation, where $\chi_p^{(1)} = N_a|\mathbf{e}_p \cdot \mathbf{P}_{12}|^2 \rho_{31}^{(1)}/(\varepsilon_0\hbar)$ is the linear susceptibility corresponding to the moiré lattice, $\chi_{p,1}^{(3)} = N_a|\mathbf{e}_p \cdot \mathbf{P}_{12}|^2 \rho_{31,1}^{(3)}/(\varepsilon_0\hbar)$ and $\chi_{p,2}^{(3)}(\mathbf{r}'-\mathbf{r}) = N_a^2|\mathbf{e}_p \cdot \mathbf{P}_{12}|^2 \rho_{31,2}^{(3)}V(\mathbf{r}'-\mathbf{r})/(\varepsilon_0\hbar)$ are local and nonlocal cubic nonlinear susceptibilities, respectively. Here $\varepsilon_0$ is the vacuum dielectric constant, $\chi_{p,1}^{(3)}$ depends on the interaction between light and atoms, and $\chi_{p,2}^{(3)}$ is determined by the RRI [19]. From these expressions above, one can see that $\chi_p^{(1)}$ and $\chi_{p,1}^{(3)}$ have a linear dependence on the atomic density $N_a$, $\chi_{p,2}^{(3)}$ has a quatrotic dependence on the atomic density ($N_a^2$). Further studies have shown that the fifth-order nonlinear optical susceptibilities has cubic dependence on the atomic density ($N_a^3$)[22]. This implies that the nonlinear optical susceptibilities in the Rydberg-EIT system are very sensitive to the change of the atomic density $N_a$. There exists a competition between the

third- and fifth-order nonlinear optical susceptibilities when $N_a$ becomes larger. In the third-order approximation, the atomic density should not be larger than the order of $10^{10}$.

When the probe field is weak, a perturbation expansion can be employed to solve the above equation. Under the action of the Rydberg-Rydberg interaction, the one-body correlations $\rho_{ij}(\mathbf{r},t) = \langle \hat{S}_{ij}(\mathbf{r},t) \rangle$ are affected by two-body $\rho_{33,3i}(\mathbf{r}',\mathbf{r},t) = \langle \hat{S}_{33}(\mathbf{r}',t)\hat{S}_{3i}(\mathbf{r},t) \rangle$, three-body $\rho_{ij,\mu\nu,\delta\kappa}(\mathbf{r}'',\mathbf{r}',\mathbf{r},t) = \langle \hat{S}_{ij}(\mathbf{r}'',t)\hat{S}_{\mu\nu}(\mathbf{r}',t)\hat{S}_{\delta\kappa}(\mathbf{r},t) \rangle$, and higher-order correlations. The detailed derivation of two- and three-body correlations can be found in Ref [25]. The physical quantities are expanded in powers of a small amplitude $\varsigma$ as: $\Omega_p = \sum_{m=1} \varsigma^m \Omega_p^{(m)}$, $\rho_{j1} = \sum_{m=0} \varsigma^{2m+1} \rho_{j1}^{(2m+1)}$, $\rho_{jl} = \sum_{m=1} \varsigma^{2m} \rho_{jl}^{(2m)}$, $\rho_{11} = 1 + \sum_{m=1} \varsigma^{2m} \rho_{11}^{(2m)}$, $\rho_{j1,l1} = \sum_{m=1} \varsigma^{2m} \rho_{j1,l1}^{(2m)}$, $\rho_{j1,1l} = \sum_{m=1} \varsigma^{2m} \rho_{j1,1l}^{(2m)}$, $\rho_{jl,\mu1} = \sum_{m=1} \varsigma^{2m+1} \rho_{jl,\mu1}^{(2m+1)}$, and $\rho_{jl,\mu\nu} = \sum_{m=2} \varsigma^{2m} \rho_{jl,\mu\nu}^{(2m)}$ $(j,l,\mu,\nu = 2,3,4,5)$. Accordingly, scaled coordinates are defined as $x_1 = \varsigma x$, $y_1 = \varsigma y$, $z_l = \varsigma^l z (l=0,1,2)$, and $t_l = \varsigma^l t (l=0,1)$. Substituting these expressions into Maxwell's equation and Bloch's equation, the solution for the probe fields can be obtained order by order[25].

Thus the effective nonlocal nonlinear Schrodinger equation (NNLSE) for the probe field can be derived in a normalized form:

$$i\frac{\partial \psi}{\partial s} + \left(\frac{\partial^2 \psi}{\partial \xi^2} + \frac{\partial^2 \psi}{\partial \eta^2} + \frac{\partial^2 \psi}{\partial \tau^2}\right) - V_{\text{OL}}\psi + \gamma |\psi|^2 \psi + \alpha \int d^3\mathbf{r}' V(\mathbf{r}' - \mathbf{r})|\psi|^2 \psi = 0 \quad (7)$$

where $\psi = \Omega_p/\psi_0$, $s = z/(2L_{\text{diff}})$, $\tau \to (t - z/v_g)/\tau_0$, $(\xi, \eta) = (x, y)/R_\perp$, with $\psi_0$, $L_{\text{diff}} = \omega_p R_\perp^2/c$, $R_\perp$, $v_g = (\partial K/\partial \omega)^{-1}$ and $\tau_0$ being the initial Rabi frequency of the probe pulse, diffraction length, typical radius of the probe pulse, group velocity, and pulse duration, respectively. Further $\gamma$ is the local-nonlinearity coefficient representing the repulsive inter-atomic interactions. $\alpha$ is the strength of the nonlocal nonlinearity, which accounts for the effective long-range attractive interaction between Rydberg atoms, and can be characterized as $\alpha = R_b/R_\perp$ with the local region($\alpha \ll 1$), general nonlocal region($\alpha \sim 1$), and strongly nonlocal region($\alpha \gg 1$)[19]. Here the Rydberg blockade radius is obtained as

$R_{b}=\left[\left|C_{6}\Delta_{3}\right|/\left|\Omega_{c}\right|^{2}\right]^{1/6}$. Note that the nonlocal nonlinearity $\alpha$ can be driven by parameters, such as $\Omega_{c}$ and $R_{\perp}$. NNLSE (7) is the basic equation used in this work. Its relevant solutions, presented below, were produced in numerical form.

If the nonlocal nonlinearity is dropped in Eq. (7), setting $\alpha=0$, the remaining equation with the local self-defocusing (represented by $\gamma<0$ still gives rise to gap solitons (GSs, including those with embedded vorticity). Then, it is necessary to explicitly stress what is essentially new in the existence and stability of GSs in the model including the nonlocal term, in comparison to the GSs existing in the absence of the nonlocal terms. In this work, due to the presence of the long-range interactions between Rydberg atoms, the combination of the local repulsive and nonlocal attractive nonlinearities in the laser-coupled atomic system, together with the action of ML-induced potential, makes it possible to create complex but, nevertheless, stable static and dynamical self-trapped topological modes, as shown below.

For experimental consideration, this model can be realized, in particular, in a cold gas of $^{88}$Sr atoms[19]. In that case, relevant values of the physical parameters are $R_{\perp}=12\mu m$, $\tau_{0}=1.1\times10^{-6}s$, $\Gamma_{21}=0.2\pi MHz$, $\Gamma_{3}=2\pi\times16MHz$, $\Gamma_{4}=\Gamma_{34}=2\pi\times16.7kHz$, $\Delta_{2}=1.67\times10^{6}s^{-1}$, $\Delta_{3}=9.67\times10^{7}s^{-1}$, $\Delta_{4}=1.36\times10^{7}s^{-1}$, $N_{a}=1.0\times10^{10}cm^{-3}$, $\Omega_{c}=\Omega_{c0}=1.2\times10^{7}s^{-1}$, $\Omega_{a}=\Omega_{a0}=5\times10^{6}s^{-1}$. The integration is regularized by restricting it to $|\mathbf{r}'-\mathbf{r}|\geq10^{-4}mm$, which corresponds to the extremely large size of Rydberg atoms. The unit of the accordingly scaled propagation distance, $s=1$, corresponds to $\simeq1$ mm.

**3. Numerical results**

We now discuss the formation of LBs in the system. Stationary solutions of Eq. (7) can be sought as

$$\psi=\varphi(\mathbf{r})e^{ibs}, \qquad (8)$$

where $b$ is propagation constant. The stationary solution $\varphi(\mathbf{r})$ can be sought as $\varphi(\mathbf{r})=\phi_{\mathbf{k}}(\mathbf{r})\exp(i\mathbf{k}\cdot\mathbf{r})$, where the wave vectors $\mathbf{k}=(k_{\xi},k_{\eta})$ are confined to the first reduced Brillouin zone of the moiré lattices. The band structure about $b$ can be found by calculating the linear eigenvalue problem with plane wave expansion method[33,41], i.e.,

$$\left[\frac{1}{2}\left(\frac{\partial}{\partial \mathbf{r}}+i\mathbf{k}\right)^2 - V_{OL}(\mathbf{r})\right]\phi_{\mathbf{k}}(\mathbf{r}) = b\phi_{\mathbf{k}}(\mathbf{r}). \tag{9}$$

The numerical scheme is based on the following procedure: stationary solutions of different types are obtained through the modified squared-operator iteration method [42], and direct simulations of Eq. (7) are performed using the fourth-order Runge-Kutta method. $s$ is the propagation distance with step size $\Delta s = 0.01$ in the direct simulations of the evolution. The computations were performed on the grid with mesh size $10^{-4}$, and the boundary conditions are $|\psi|^2 = 0$ at $\eta = \pm 4$, $\xi = \pm 4$ and $\tau = \pm 2$.

The stability of the stationary solutions was identified, first, through the linear-stability analysis. To this end, perturbed solutions for the wave function were looked for as $\psi = (\varphi + pe^{\lambda s} + q^* e^{\lambda^* s})e^{ibs}$, where $\varphi$ is the stationary solution (7), while $p$ and $q$ represent an eigenmode of small perturbations with instability growth rate $\lambda$. The soliton solutions are stable, as usual, if all eigenvalues have $\text{Re}(\lambda) = 0$. Then, to test the stability or instability in direct numerical simulations, we added small-amplitude white noise to the input at $t = 0$ for stable solitons, and did not add noise to unstable ones, as in the latter case small errors of the numerical scheme were sufficient to trigger the growth of the instability. The energy of the probe field is defined as $U = \int |\psi|^2 \, d\xi d\eta d\tau$ in the direct simulations.

The necessary condition for the stability of LBs supported by a self-defocusing nonlinearity is provided by the anti-Vakhitov-Kolokolov (anti-VK) criterion $dU/db > 0$ [43], and full stability is determined by the spectrum of eigenvalues $\lambda$ for small perturbations, produced by Eq. (7) linearized around the stationary state.

In this work, we construct zero-vorticity and vortical GS families. In addition to that, we produce two species of excited states of the zero-vorticities solitons,of dipole and quadrupole types.

### 3.1 Zero-vorticity GSs

Our analysis reveals three species of zero-vorticity LBs, *viz*., fundamental, dipole and quadrupole ones. The dipole and quadrupole complexes have opposite signs of the wave field in adjacent components. Parameters of Eq. (7), such as the nonlinearity coefficients $\gamma$ and $\alpha$, and ML parameters $(\theta, V_1, V_2)$, are tuned to produce stable LBs. Typical examples and proof of their stability in direct simulations are displayed in Fig. 2. The profiles are shown for four propagation distances, $s = 10, 100, 300, 500$.

In the course of propagation, the LBs keep their stability. Some fuzziness observed at $s=500$ in Fig. 2 eventually emerges from a small numerical inaccuracy of the numerically found stationary solutions, which are used as inputs in the direct simulations. It is relevant to mention that this feature becomes weakly visible after the propagation distance which is tantamount to ~500 diffraction (Rayleigh), which, in that sense, is a very long distance, implying that the solitons are well-established objects. This work shows robust LBs in the simulation and provides a way for researching solitons experimentally.

For the dipole LBs, the distance between the centers of the two components keeps value $d=7.68$, while the distances between components of the quadruple are kept as $d_1=d_3=4.73$ and $d_2=6.68$.

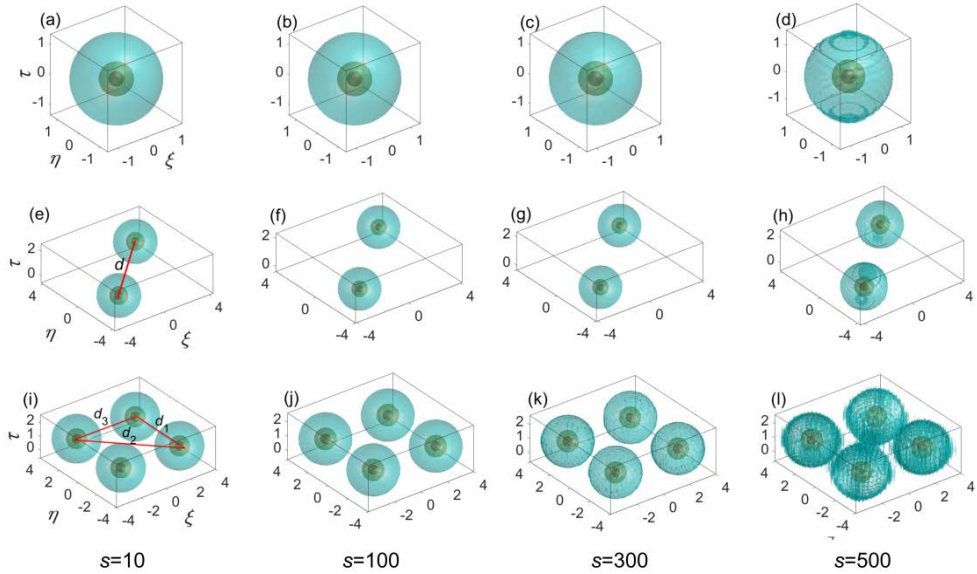

Fig. 2 Isosurfaces of stable gap solitons at different propagation distances, $s=10,100,300,500$, as produced by simulations of Eq. (7). Parameters are $\alpha=0.5$, $\gamma=0.5$, $\theta=1$, $V_1=V_2=1$ for the fundamental LB (a-d); $\alpha=0.5$, $\gamma=0.15$, $\theta/\pi=0.76$, $V_1=V_2=1.2$ for the dipole LB (e-h); $\alpha=0.9$, $\gamma=0.2$, $\theta/\pi=0.8$, $V_1=V_2=10$ for the quadrupole LB(i-l). The lattice strength is fixed as $V_0=1$. In panels (e) and (i), $d$ and $d_{1/2/3}$ are distances between centers of separate components of the dipole and quadrupole LBs. The isosurfaces correspond to fixed values of the absolute value of the wave field, $|\psi(\xi,\eta,\tau)|=(0.90,0.53,0.05)|\psi|_{\max}$.

The shape, energy, and stability of the LBs can be controlled by varying the system's parameters. Dependences of the LBs' energy $U$ on parameters $\alpha$, $\gamma$, $p$ and $\theta/\pi$ are shown in Fig. 3, where point A designates the stable LBs from Fig. 2. For fundamental LBs, $U$ decreases monotonously with the increase of the nonlinearity coefficients $\gamma$ and $\alpha$. Similar results are observed for the dipole and quadrupole LBs. Stability zones corresponding to Fig. 3 and Fig. 4 are: $0.1 < b < 8.4$, $0.1 < \alpha < 1.2$, and $0 < \gamma < 1.1$ for the fundamental LBs; $0 < b < 6.2$, $0.1 < \alpha < 0.7$, and $0.1 < \gamma < 0.5$ for the dipole LBs, $3 < b < 8.98$, $0.26 < \alpha < 1.0$, and $0 < \gamma < 0.58$ for the quadrupole LBs (the stability intervals for $b$ correspond to Fig.4.

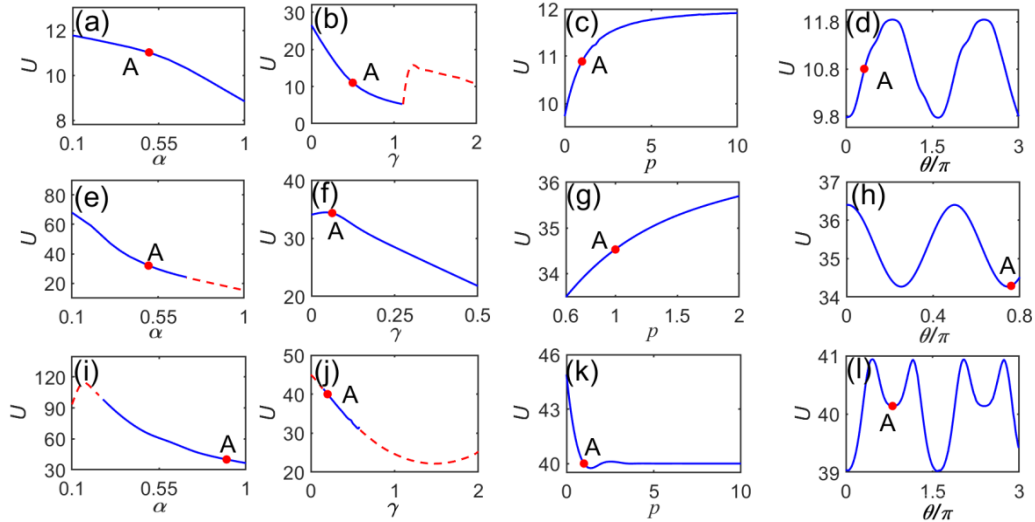

Fig. 3 The LB's energy $U$ as functions of the system parameters, $\gamma$, $\alpha$, $p$ and $\theta/\pi$ for LBs of the fundamental (a-d), dipole(e-h), and quadrupole (i-l) types. Point A in these subplots represents the stable LBs shown in Fig.2. Solid/dashed lines represent stable/unstable states. Parameters that are not varied in the plots are the same as Fig. 2.

Unlike the dependences on the nonlinearity coefficients $\gamma$ and $\alpha$, a relatively weak dependence of energy $U$ is observed on the ML parameters, $p$ and $\theta$, as shown in Fig. 3(c,d,g,h,k,l). In most cases, relative changes of $U$ in these panels are smaller than 1%. The only exception is a fast decrease of $U$ with the increase of $p$ in Fig. 3(k) at $P$.

Because the ML induces the spatially periodic potential for the atomic gas, energy $U$ displays periodic variation, with a period $T_\theta = 1.6\pi$, as a function of the ML's twist angle $\theta$. For the dipole LBs, the dependence is smoother, and the period is smaller, $T_\theta = 0.5\pi$. The period for the quadrupole LBs is the same as for the dipole, $T_\theta = 1.6\pi$, with two peaks per period. The reason is that spatially separated components of the quadrupole partly overlap in the 3D space.

In Figs. 4(a,d,g), we display $U(b)$ dependencies for the fundamental, dipole and quadrupole LBs, respectively. In particular, it is observed that the anti-VK criterion, $dU/db > 0$, is indeed a necessary stability criterion.

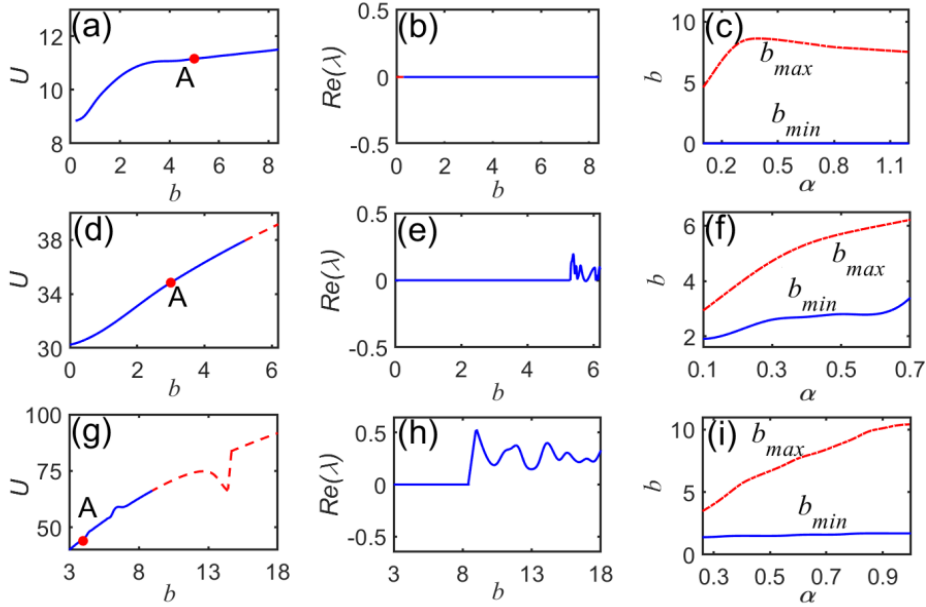

Fig. 4. The left and central columns show the dependence of energy $U$ and the real part of stability eigenvalue (instability growth rate) $\lambda$ on the propagation constant, $b$, for the fundamental (a,b), dipole (d,e), and quadruple (g,h) LBs, respectively. Point A represents the stable LBs from Fig.2. The solid/dash lines represent the stable/unstable states. Panels (c,f,i) show boundaries of the stability intervals, $b_{min} < b < b_{max}$, as a function of strength $\alpha$ of the nonlocal term in Eq (7) for the LBs of the fundamental, dipole, and quadrupole types, respectively. Parameters that are not varied in the plots are the same as Fig. 2.

As mentioned above, full stability of the LBs is determined by the real part of eigenvalues $\lambda$ for small perturbations, $\mathrm{Re}(\lambda)$, the solitons being stable if $\mathrm{Re}(\lambda) = 0$. Dependences of $\mathrm{Re}(\lambda)$ on the propagation constant $b$ for the fundamental, dipole and quadrupole LBs are shown in Figs. 4(b,e,h), respectively. It is seen that the family of the fundamental LBs is completely stable, while the increase of $b$ leads, eventually, to the onset of instability in the dipole and quadrupole families.

For fixed parameters of Eq. (7), the LBs are stable in intervals $b_{\min} < b < b_{\max}$. The dependence of the edges of the intervals, $b_{\min}$ and $b_{\max}$, on the nonlocal-nonlinearity coefficient $\alpha$ in Eq. (7) are shown in Figs. 4(c,f,i) for the fundamental, dipole and quadrupole LB families, respectively. It is seen that the stability zone is much larger for the fundamental LBs, which actually have $b_{\min} = 0$, i.e., this family is bounded only from above, $b < b_{\max}$. In other systems, the stability interval of the propagation constant for various modes may also be bounded from below by a certain minimum value [44-46].

The widths of LBs in the spatial and temporal directions are defined as $W_\xi^2 = \int \xi^2 |\psi|^2 \mathrm{d}\xi \mathrm{d}\eta \mathrm{d}\tau / U$ and $W_\tau^2 = \int \tau^2 |\psi|^2 \mathrm{d}\xi \mathrm{d}\eta \mathrm{d}\tau / U$, where $U = \int |\psi|^2 \mathrm{d}\xi \mathrm{d}\eta \mathrm{d}\tau$. The transverse integration is defined with respect to scaled spatial coordinates $\xi$, $\eta$ and reduced time $\tau$. Further, the LB's overall width is defined as

$$W(s) = \int |\psi| \mathrm{d}\xi \mathrm{d}\eta \mathrm{d}\tau / \sqrt{U}. \tag{10}$$

In Fig. 5, profiles of the LB states are characterized by the dependences of the shape parameters $W_\xi$, $W_\tau$, and $d_i$ on the control parameters, $\alpha$, $\gamma$ and $P$. For the spherically isotropic fundamental LBs, the spatial and temporal sizes are equal in the present notation, $W_\xi = W_\tau$, as seen in Fig. 5(a-c). The variation of the width is small ($<10\%$) in the stability zone, and the width does not vary following the change the ML parameter $P$. In addition to that, Fig. 5(d) shows that fluctuations of the scaled width of the fundamental LB, $R=W(s)/W(0)$, remain very small in the course of the propagation, and its shape keeps the spherical isotropy. Thus the fundamental LBs are more easier to generate and can stably propagate without deformation.

The shapes of the dipole and quadrupole LBs do not vary much either, following the change of the control parameters [see Figs. 5(e-g) and (i-k), respectively]. However, unlike the fundamental LBs, their spatial and temporal widths are not equal, $W_\xi \neq W_\tau$, as the dipoles and quadrupoles are not spherically isotropic states. As a result, the width ratio $W_\xi/W_\tau$ varies between 2 and 3.

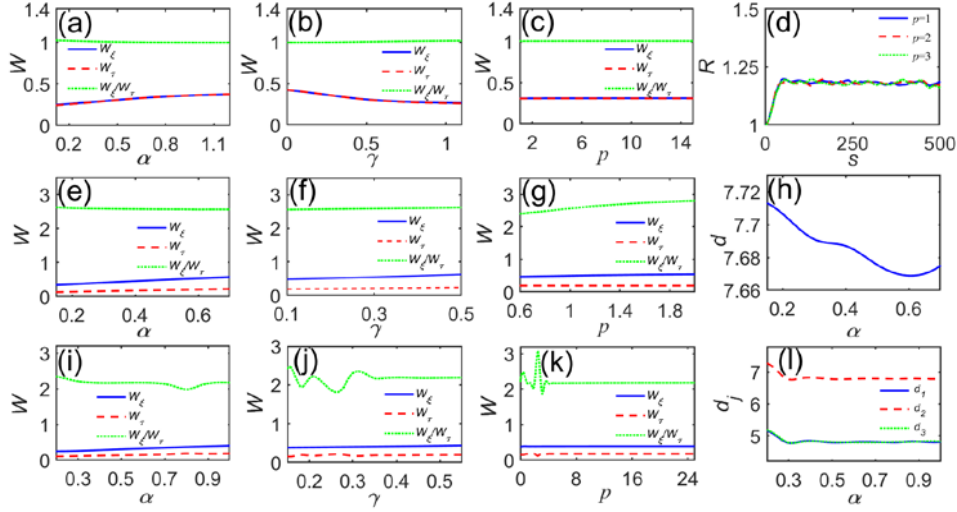

Fig. 5. The LBs' widths $W$ and distances $d$ between components of the dipoles and quadrupoles vs. parameters $\alpha$, $\gamma$ and $p$ of Eq. (7). Panels (a-d), (e-h), and (i-l) pertain to the fundamental, dipole, and quadrupole LBs, respectively. Panel (d) additionally shows the variation of the scaled width of the fundamental LB, $R=W(s)/W(0)$, in the course of the propagation.

For the dipole LBs, the relative variation of the separation between their components is small too, $\Delta d/d < 1\%$ in the stability interval of the nonlocal-nonlinearity coefficient $\alpha$, as shown in Fig. 5(h). For the quadruple LBs, separations $d_{1/2/3}$ decrease with the growth of $\alpha$ at $\alpha < 0.3$, and slightly fluctuate at larger values of $\alpha$, see Fig. 5(l).

The conclusion is that the zero-vorticity LBs of the fundamental, dipole, and quadrupole types are robust states. This fact is beneficial for use of these LB modes in photonic schemes.

## 3.2 Vortex GSs

The search for vortex solitons was initiated by an initial guess

$$\psi = Ar\exp(-Br^2 - Cz^2 + iS\varphi), \tag{11}$$

where $r$, $z$ and $\varphi$ are the cylindrical coordinates, $A$, $B>0$ and $C>0$ are constants, and $S$ is the integer vorticity. In our simulations, we consider the case of $S=1$, as the numerical solutions showed that the balance with coexistence of moiré lattice potential and Rydberg long range interaction is hard to achieve with $S>1$. This phenomenon was also found By Huang et.al. [19] that higher-order vortices may not keep their shape in the nonlocal response region of cold Rydberg gases.

Isosurfaces for stable and unstable vortex LBs are displayed in Fig. 6. In the course of the propagation, the stable one, with propagation constant $b=1.8$, maintains its structure. The vortex state corresponding to $b=2$ is completely unstable, quickly splitting into a necklace-shaped set of four fragments. The emerging necklace may be classified as a quasi-stable state, as it keeps its shape, while the verification of its full stability requires extremely long simulations, corresponding to propagation distances that may exceed those achievable in the experiment.

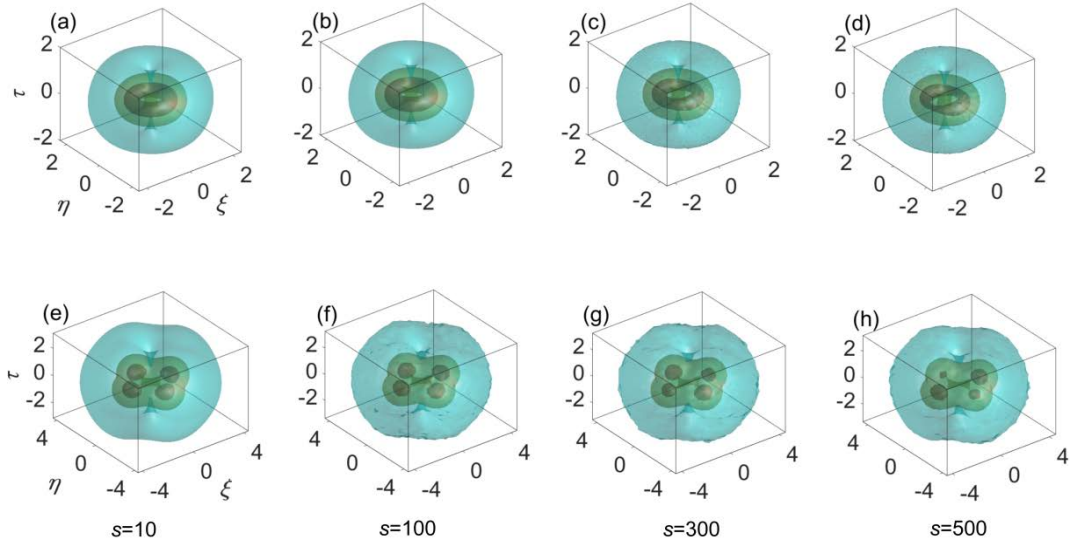

Fig. 6 Isosurfaces of vortex GSs at different propagation distances $s=10,100,200,300$. Stable (a-d) and unstable (e-h) vortex LBs are obtained with $b=1.8$ and $b=2$, respectively. Other parameters are $\gamma=0.7$, $\alpha=0.7$, $\theta/\pi=0.5$, and $V_1=V_2=1$.

It is known that the anti-VK criterion does not detect instability of vortex soliton against spontaneous splitting [6,43], therefore their stability should be explored through the computations of eigenvalues for small perturbations, as shown in Fig. 7. As seen in Figs. 7(a,b), the vortex LBs are stable in a finite interval of values of the propagation constant, $b_{min} \leq b \leq b_{max}$. The dependence of $b_{min}$ and $b_{max}$ on the nonlocal-nonlinearity coefficient $\alpha$ is presented in Fig. 7(c), where, in particular, $b_{min} = 0$, similar to the situation for the fundamental zero-vorticity LBs, cf. Fig. 4(c).

Dependences of light energy $U$ of the vortex LBs on the control parameters, $\alpha$, $\gamma$, $p$ and $\theta$, are displayed in Figs. 7(d-f), which shows that the vortex-LB family is chiefly a stable one. It is observed that $U$ decrease monotonously with the growth of nonlinearity coefficients $\gamma$ and $\alpha$, while the dependence on the ML parameters, $p$ and $\theta$, is weak. Similar to the fundamental LBs [cf. Figs. 3(d,h,l)], the energy of the vortex LBs is a periodic function of the twist angle $\theta$, although the width of the periodic variation range is very small. The stability zones for the vortex LBs are $0.3 < b < 2.6$, $0.6 < \alpha < 1.5$ and $0.4 < \gamma < 1.5$.

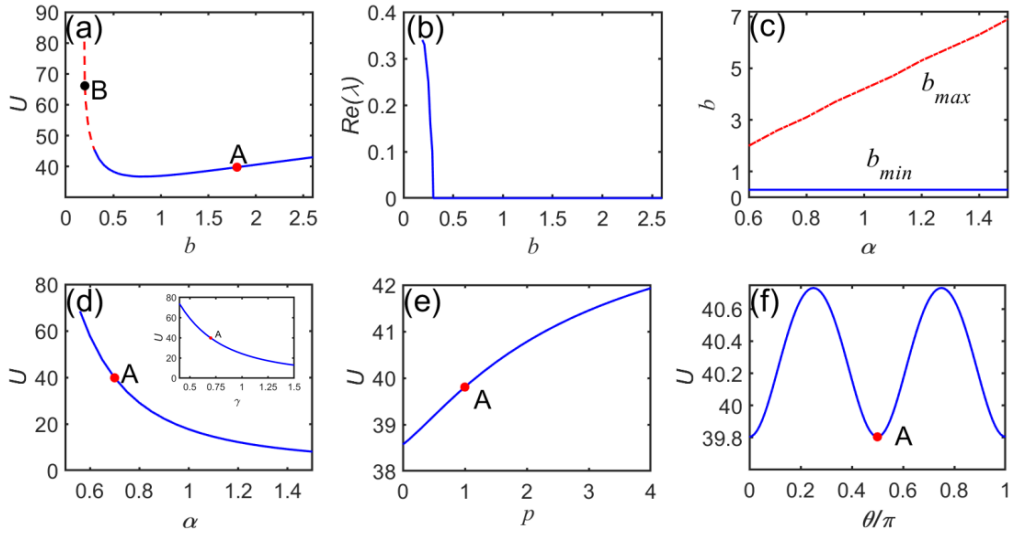

Fig. 7 (a) The energy of the vortex LBs vs. propagation constant $b$. (b) The real part of the stability eigenvalue $\text{Re}(\lambda)$ (i.e., the instability growth rate) vs. $b$. (c) Boundaries of the stability interval, $b_{min} \leq b \leq b_{max}$, vs. $\alpha$. (d,e,f) The energy of the stable vortex LBs vs. parameters $\alpha$, $\gamma$, $p$ and $\theta$ [the dependence on $\gamma$ is shown by the inset in panel (d)]. Points A and B correspond to the stable and unstable states in Fig. 6. The parameters which are not varied are the same as Fig. 6. The vortex-LB families shown in panels (d)-(f) are entirely stable.

## 4. Conclusion

We have introduced the four-level cold Rydberg atomic system with the spatially periodic and quasi-periodic optical potentials induced by the MLs (moiré lattices). The Rydberg-dressed $\Lambda$-type EIT is realized in this setup, using the coupling of the probe, control, and auxiliary laser fields to the four atomic levels. The system includes the local defocusing and nonlocal attractive cubic nonlinearities, induced by the Rydberg interactions. The stationary solutions of different types are obtained through the modified squared-operator iteration method and direct simulations on grids with mesh size $10^{-4}$. The system's bandgap spectrum is found, and three-dimensional LBs (light bullets) are produced in the form of GSs (gap solitons). By tuning the control parameters, such as the nonlinearity coefficients and ML parameters, LB families of the fundamental, dipole, quadruple, and vortex types are produced. Stability regions for them are investigated using the anti-VK criterion and computation of eigenvalues for small perturbations. The light energy and shapes of the LBs can be effectively by tuning the nonlinearity coefficients and ML parameters, while the dependence on the ML's twist angle is periodic. The results suggest new options for manipulating the 3D nonlinear optical modes in Rydberg-gas media. Future research interests may be the vortex solitons with higher topological charges in nonlocal nonlinear systems with Rydberg atoms at twist coordinates system.

**Funding**

The work of S.L. Xu was supported by the National Natural Science Foundation of China (62275075). The work of Y. Zhao was supported by the Natural Science Foundation of Hubei province (2023AFC042). The work of B.A.M. was supported, in part, by the Israel Science Foundation through grant No. 1695/22.

**Conflict of Interest Disclosure**

The authors declare no competing financial interest.